# Single-Crystal Organic Charge-Transfer Interfaces probed using Schottky-Gated Heterostructures

Ignacio Gutiérrez Lezama[1], Masaki Nakano[2], Nikolas A. Minder[1], Zhihua Chen[3,4], Flavia V. Di Girolamo[1,5], Antonio Facchetti[3,4] and Alberto F. Morpurgo[1*]

[1] DPMC and GAP. University of Geneva, 24 quai Ernest Ansermet, CH1211 Geneva, Switzerland

[2] CERG, RIKEN Advanced Science Institute, Wako, Saitama 351-0198, Japan

[3] Polyera Corporation, 8045 Lamon Avenue, Skokie, Illinois 60077, USA

[4] CHEM, Northwestern University, 2145 Sheridan Road, Evanston, Illinois 60208, USA

[5] CNR-SPIN and University of Naples, p.le Tecchio 80 80125 Naples, Italy

* To whom correspondence should be addressed. E-mail: alberto.morpurgo@unige.ch

**Organic semiconductors based on small conjugated molecules generally behave as insulators when undoped, but the hetero-interfaces of two such materials can show electrical conductivity as large as in a metal. Although charge transfer is commonly invoked to explain the phenomenon, the details of the process and the nature of the interfacial charge carriers remain largely unexplored. Here we use Schottky-gated heterostructures to probe the conducting layer at the interface between rubrene and PDIF-CN$_2$ single crystals. Gate-modulated conductivity measurements demonstrate that interfacial transport is due to electrons, whose mobility exhibits band-like behavior from room temperature to ~ 150 K, and remains as high as ~ 1 cm$^2$V$^{-1}$s$^{-1}$ at 30 K for the best devices. The electron density decreases linearly with decreasing temperature, an observation that can be explained quantitatively based on the heterostructure band-diagram. These results elucidate the electronic structure of rubrene-PDIF-CN$_2$ interfaces and show the potential of Schottky-gated organic heterostructures for the investigation of transport in molecular semiconductors.**

Organic semiconductor interfaces determine the operation of devices based on molecular materials[1], such as organic light emitting diodes[2,3], *p-n* junctions[4,5], and solar cells[6,7,8]. In most cases, electronic transport perpendicular to the interface is the key process of interest. Transport in the interface plane has attracted attention only recently[9,10,11,12,13,14,15,16,17,18] (for early work see ref. 9), both in the applied[9,10,12,13,14] and fundamental domains[15]. It has been found that interfaces between two organic semiconductors regularly exhibit an enhanced electrical conductivity, and that, when single-crystals of specific molecules are used, this interfacial conductivity can be as large as in a metal –a remarkable observation, given the insulating character of the constituent materials. There is consensus that charge transfer from the surface of one organic semiconductor to that of the other determines the interfacial conductivity [15,16,19], but even the most basic properties of these organic "charge-transfer" interfaces could not be accessed experimentally so far, preventing a true understanding of the phenomenon. Indeed, interpretation of experiments has so far relied on plausible physical scenarios and estimates[15,16], but no direct measurements could be performed to determine whether the interfacial charge carriers are electrons or holes (or both), what is their density, their mobility, how these quantities change as a function of temperature, or whether they can be tuned in device structures.

Here we address these problems by performing experiments on a new type of device: organic Schottky-gated heterostructures based on organic single crystals. The devices that we discuss (see Fig. 1a-e) are based on rubrene[20] and PDIF-CN$_2$ single crystals[21] –the materials that perform best in *p*- and *n*-channel organic single-crystal field-effect transistors (FETs)- at whose interface a highly conducting layer forms spontaneously (Fig. S1). The heterostructures consist of a Cr film (the Schottky gate) deposited on a PDMS support, onto which first a rubrene and then a PDIF-CN$_2$ single crystal are laminated. Before laminating the PDIF-CN$_2$ crystal, gold electrodes are evaporated onto rubrene through a shadow mask, to contact the interfacial conducting layer. A schematic diagram and actual images of devices are shown in Fig. 1 (the details on the fabrication process are discussed in the supplementary information). Devices of this type have never been previously realized with organic semiconductors, but have clear analogies with Schottky-gated heterostructures based on III-V inorganic semiconductors[22].

The Schottky barrier present at the Cr-rubrene interface prevents charge injection for one polarity of applied bias, effectively isolating the gate electrode from the rubrene-PDIF-CN$_2$ interface[23, 24]. Owing to the depletion region associated to the Schottky barrier and the very low density of states in the band-gap of rubrene single-crystals, virtually all charge induced upon application of a gate voltage is accumulated at the rubrene-PDIF-CN$_2$ interface. This eliminates problems present in heterostructures with a "conventional" gate –i.e., a gate separated from the semiconductor by an insulating layer- in which the induced charge is also accumulated at the interface between the gate insulator and the semiconductor (see for example ref. 19). As it is important to both ensure correct device operation and characterize the heterostructures –and given that Schottky gates are not commonly employed in conjunction with organic semiconductors- we start by discussing measurements showing that the gate electrode indeed functions as expected for a Schottky contact.

Fig. 2a shows the current flowing through the rubrene crystal upon biasing the gate (i.e., the current between the gate and both the drain and source contacts; during device operation this corresponds to the gate leakage current). This current depends on the bias polarity, as expected for a Schottky gate. For negative gate voltages (see Fig. 2b), the gate extracts holes injected at the rubrene-PDIF-CN$_2$ interface, giving rise to a space charge limited current[25] through the rubrene crystal. Accordingly, the *I-V* curves (Fig. 2c) show an approximately quadratic bias dependence. For positive gate voltages the current is much smaller (see Fig. 2b), since –as mentioned above- holes injected from the chromium gate into rubrene have to overcome the Schottky barrier (the work function of chromium is ~4.5 eV[23], corresponding to a Schottky barrier height of approximately 0.8 eV[24, 26]). As the temperature is reduced, the leakage current decreases following a thermally activated mechanism. For positive gate bias this is due to thermal activation over the barrier, while for negative bias (Fig 2d) the thermal activated behavior originates from space charge limited current through states in the band-gap of rubrene (i.e., through traps[27, 28]). The activation energy (inset of Fig. 2d), corresponding roughly to the energy difference between the Fermi level and the bottom of the valence band of rubrene, ranges in between 250 and 400 meV, in agreement with recent studies of rubrene metal-semiconductor field-effect transistors[24] and previous investigations of space-charge limited current[28]. The observed variations are indicative of sample-to-sample fluctuations in the concentration of dopants unintentionally present in rubrene and in the density of in-gap states, which determine the position of the Fermi level relative to the top of the valence band.

Having ensured the proper operation of the Schottky gate, we proceed to discuss transport in the interface plane, which was investigated as a function of source-drain bias ($V_{DS}$), gate voltage ($V_G$), and temperature ($T$). Fig. 3a shows that, even at $V_G = 0$ V, the $I$-$V$ curves of the heterostructures are linear throughout the investigated temperature range, and that the resistance is reproducibly close to 1 M$\Omega$/□ at room temperature (see the table in the supplementary information for a summary of data from all working devices). This observation directly indicates that a highly conducting layer does indeed form at the rubrene-PDIF-CN$_2$ interface, since the resistance of individual rubrene and PDIF-CN$_2$ crystals is much larger than 1 G$\Omega$. The resistance (Fig 3b) exhibits only rather small variations when $T$ is reduced down to ~100-150 K, whereas further lowering causes a more pronounced resistance increase for all devices. Fig. 3c and d show the gate-voltage dependence of the current flowing in these heterostructure devices. The measured current always increases with applying a positive voltage to the Schottky gate and decreases for the opposite polarity, exhibiting a nearly linear $V_G$-dependence throughout the investigated range. This observation indicates an $n$-type behavior of the devices, from which we conclude that the current in the interfacial conducting layer is carried by electrons.

We estimate both the electron density and field-effect mobility using the same type of analysis commonly adopted for conventional FETs. The interfacial conductivity can be written as $\sigma(V_G) = n(V_G)e\mu_{FET}$ where $n(V_G)$ is the density of electrons at the rubrene-PDIF-CN$_2$ interface, given by $n(V_G) = C(V_G-V_T)/e$ ($C = \varepsilon\varepsilon_0/d$ is the capacitance per unit area between the interface and the gate, $\varepsilon$ is the relative dielectric constant of rubrene and $d$ is the thickness of the rubrene crystal –close to 2 μm in most devices; $V_T$ is the threshold voltage, corresponding to the extrapolated gate voltage at which the source-drain current vanishes). The mobility can therefore be extracted as $\mu_{FET} = L/W \; 1/CV_{DS} \; dI_{DS}/dV_G$ (where $L$ and $W$ are the interface length and width, respectively) and is found to be nearly gate-voltage independent. The electron density as a function of $V_G$ is then given by $n = \sigma/e\mu_{FET}$, where $\sigma$ is the conductivity.

The temperature dependence of the field-effect mobility and electron density at $V_G = 0$ for three devices are shown in Fig. 4a and b. The room temperature electron mobility values compare well to those found for FETs based on PDIF-CN$_2$ single-crystals with Cytop or PMMA as the gate dielectric[29], which have approximately the same dielectric constant of rubrene[30, 31, 32]. Consistency in the mobility values supports the validity of our analysis and confirms that the electrons are accumulated at the rubrene-PDIF-CN$_2$ interface. It also indicates that the surfaces of the PDIF-CN$_2$ crystals maintain their integrity, i.e. the crystal surface is not damaged by the proximity of rubrene, as it could occur by molecular inter-diffusion. Interestingly, as the temperature is lowered, the electron mobility exhibits a more pronounced band-like transport[20, 33] behavior as compared to PDIF-CN$_2$ single crystal FETs with suspended channel (i.e., with vacuum as gate dielectric), peaking at temperatures ~150-170 K ($\mu_{FET}$ values up to ~ 4-5 cm$^2$/Vs are observed) and remaining close to 1 cm$^2$V$^{-1}$s$^{-1}$ in the best devices even at temperatures as low as 30 K. This observation shows that rubrene-PDIF-CN$_2$ Schottky-gated devices lead to significantly improved low-temperature electron transport. The electron density $n$ is found to range between 1 x 10$^{12}$ cm$^{-2}$ and ~ 5 x 10$^{12}$ cm$^{-2}$ in different devices, and in all cases it decreases linearly as temperature is decreased, with only small (< 10%) device-to-device deviations in the slope. The high reproducibility of the linear temperature dependence and of its slope are particularly remarkable, in view of the much large sample-to-sample variations in the total electron density.

To confirm the validity of the conclusions extracted from the analysis of the field-effect response, we have also performed Hall effect measurements[29] in the presence of a perpendicular magnetic field. The fabrication of devices in a Hall bar geometry is complex, as it requires a very precise alignment of the PDIF-CN$_2$ crystal to the electrodes measuring the Hall voltage. Nevertheless, we succeeded in measuring the Hall effect as a function of temperature, enabling us to extract both the density of charge carriers and the mobility. The results are shown in Fig. 4c and 4d, respectively (see the supplementary information for the raw Hall voltage data) and agree with the analysis of the field-effect data. In particular, the charge carrier density extracted from the Hall effect decreases linearly as the temperature is lowered, with the same slope extracted from the field-effect analysis (we found that in Hall bar devices the electron mobility is regularly somewhat lower, probably due to the effect of the mechanical stress in the PDIF-CN$_2$ crystals, induced by the alignment process during fabrication).

As we now proceed to show, the experimental observations can be rationalized in terms of the semiconductor heterostructure band diagram, accounting for two sources of charge accumulation at the rubrene-PDIF-CN$_2$ interface: the charge transfer across the interface from the surface of the rubrene crystal to that of the PDIF-CN$_2$, and the charge displaced across the entire heterostructure. We start by discussing the first contribution. Having observed that the interfacial carrier density decreases only linearly –and not exponentially- with decreasing $T$, implies that no gap (or at most a very small one) is present between the top of the rubrene valence band and the bottom of the PDIF-CN$_2$ conduction band (as it is the case, for instance for TMTSF-TCNQ[16] interfaces). At the same time, the density of transferred charge carriers – few times $10^{12}$ cm$^{-2}$, 100 x smaller than in TTF-TCNQ interfaces[15]– along with its linear decrease for decreasing $T$ also imply the absence of any sizable band overlap. This implies that the top of the rubrene valence band and the bottom of the PDIF-CN$_2$ conduction band are aligned at the same energy before the two materials are brought into contact (as shown in Fig. 5a). As an independent and quantitative confirmation of the validity of this conclusion, we have measured the value of the work function difference between rubrene and PDIF-CN$_2$ crystals using scanning Kelvin force probe microscopy[34]. Since the energy difference between the Fermi level and the band edge in both rubrene and PDIF-CN$_2$ can be extracted from transport measurements on individual crystals, the difference in work function between the two materials allows us to determine the energy level alignment directly. We find that the two levels are indeed aligned with a precision of a few tens of meV (see supplementary information for details).

With such an alignment of the valence and conduction band in rubrene and PDIF-CN$_2$, interfacial charge transfer occurs at any finite temperature upon establishing contact, and stops when the electrostatic potential difference between the surfaces of the two materials, generated by the transferred charge itself, is sufficiently large. In other words, the electrostatic potential difference (an interface dipole) generated by the transferred charge shifts apart the top of the rubrene valence band and the bottom of the PDIF-CN$_2$ band, effectively opening a gap proportional to the density of transferred charge $n_S$, so that the value of $n_S$ at equilibrium needs to be found self-consistently. Within the simplest approximation –identical to the one that successfully captures the order of magnitude of the conductivity at TMTSF-TCNQ interfaces[16]- we can write (the details are found in the supporting information):

$$n_S = \int_{\frac{n_S e^2 d}{2\varepsilon\varepsilon_0}}^{\infty} 2N_S e^{-\frac{E}{kT}} dE = 2N_S kT e^{-\frac{n_S e^2 d}{2\varepsilon\varepsilon_0 kT}} \qquad (1)$$

where $\frac{n_S e^2 d}{\varepsilon\varepsilon_0}$ is the electrostatic potential difference between the rubrene and the PDIF-CN$_2$ surfaces, $N_S$ the density of states at the surface per unit energy (~2 10$^{15}$ cm$^{-2}$eV$^{-1}$), $\varepsilon$~3 the value for the relative dielectric constant of the organic materials used, and $d$~1nm the distance between the electrons accumulated on the PDIF-CN$_2$ surface and the holes on the rubrene surface. In Eq. (1) Boltzmann statistics is used because the occupation probability of states in the organic materials due to the transferred charge is much less than 1, owing to the large density of states (the physical picture used for this analysis is based on the model that we have developed to reproduce quantitatively band-like transport in organic single-crystal FETs of different molecules[29, 35]). By defining

$$y \equiv \frac{n_S}{2N_S kT} \qquad (2)$$

equation (1) becomes $y = e^{\frac{-e^2 d N_S}{\varepsilon\varepsilon_0} y}$, which does not depend on temperature anymore. If we denote its solution $y^*$, we directly find from equation (2) that

$$n_S = y^* 2N_S kT \qquad (3)$$

predicting a linear temperature dependence of the electron density on temperature, as observed experimentally. Equation (3) gives a quantitative estimate for the slope of the linear $T$ dependence that is in excellent agreement with the data (see the dashed-dotted lines in Fig. 4b). With the values of the parameters that we have indicated, the numerical solution of Eq. (2) gives $y^* = 0.029$.

In the absence of other contributions to the interfacial charge density, an equal amount of electrons and holes should be present at the surface of PDIF-CN$_2$ and rubrene. In the transport measurements, however, hole conduction is not observed. This is because the total carrier density $n$ at the interface also includes the –more conventional- contribution due to charge displaced across the entire structure, which ensures that the electro-chemical potential is spatially uniform everywhere[23]. A schematic energy band-diagram of a rubrene-PDIF-CN$_2$ heterostructure is shown in Fig. 5b (the charge transferred across the entire heterostructure does not change significantly the band discontinuity –gap- at the rubrene-PDIF-CN$_2$ interface, which depends on the amount of charge transferred from the surface of rubrene to that of PDIF-CN$_2$). At the chromium/rubrene interface, the top of the rubrene valence band is lower than the electrochemical potential by an amount corresponding to the height of the Schottky barrier. Deep in the PDIF-CN$_2$ crystal, the electrochemical potential is below the bottom of the conduction band by an amount determined by the unintentional dopants present in the crystal (which we measured to be 100-120 meV, estimated by the activation energy of the residual small conductivity of individual PDIF-CN$_2$ crystals). At a quantitative level, therefore, the details of the band-diagram depend on the doping level in the rubrene and PDIF-CN$_2$ crystals, as well as on the distance between the Schottky gate and the rubrene-PDIF-CN$_2$ interface (for ~2 μm thick crystals, it is comparable to the size of the depletion region, see ref. 24). Irrespective of these details, however, the net effect is to accumulate electrons at the interface (see supplementary information for more details) that fill the hole states of the rubrene surface and shift the chemical potential towards the electron side (thereby explaining the absence of hole conduction). Note how the overall band diagram of the device is analogous to that of conventional Schottky gate heterostructures based on inorganic materials[22].

The linear temperature dependence of the carrier density is a manifestation of the band alignment at the rubrene-PDIF-CN$_2$ interface, which is why the effect is highly reproducible in different samples. The total electron density, on the other hand, is determined by the (uncontrolled) concentration of dopants unintentionally present and by the distance between the interface and the Schottky gate, which explains the sample-to-sample fluctuations. The net result of these two contributions is to fix the electrochemical potential at the interface, close to the bottom of the conduction band of PDIF-CN$_2$, in the tail of states inevitably present due to disorder[20, 35, 36] (i.e., the shallow traps of PDIF-CN$_2$). As the temperature is lowered, the electron density decreases, lowering the chemical potential at the PDIF-CN$_2$ surface. At the same time the gap between the rubrene valence band and PDIF-CN$_2$ conduction band –which is proportional to the charge density transferred from one surface to the other- also decreases. As a result, the position of the chemical potential relative to the bottom of the PDIF-CN$_2$ conduction band is not much affected (i.e., it is "pinned" close to the bottom of the PDIF-CN$_2$ conduction band). This is the main mechanism for the high FET mobility values observed at 30-40 K, since the proximity of the chemical potential to the bottom of the PDIF-CN$_2$ conduction band results in a large concentration of electrons in the band (thermally activated out of the shallow traps), even at the lowest temperature of our measurements. Indeed, the data in Fig. 4a and b show that at low temperature there is a correlation between electron density and field-effect mobility, with higher densities corresponding to larger mobility values. The situation is different in PDIF-CN$_2$ FETs based on suspended single crystals –the devices that exhibit the largest room temperature mobility[29]- in which the maximum density of charge carriers that can be gate-induced is ~$10^{11}$ cm$^{-2}$. The resulting larger distance between chemical potential and bottom of the PDIF-CN$_2$ conduction band causes an exponential suppression of the number of charge carriers activated out of the shallow trap states, leading to orders-of-magnitude lower field-effect mobility at 30-40 K (Fig. 4a).

In summary, our results represent the first, detailed characterization of the electronic properties of single-crystal organic charge-transfer interfaces, enabling the determination of the nature of the interfacial charge carriers, as well as their temperature dependent density and mobility. Their theoretical analysis links successfully the observed transport properties to the interfacial electronic structure, in a way that can be cross-checked experimentally (for instance, through the Hall effect measurements and the scanning Kelvin force probe microscopy). It is not yet common, in the field of organic semiconductors, to be able to establish such a detailed and consistent description of the low-energy electronic properties of artificially realized structures. The possibility to do so in this case is largely due to the use, for the experimental investigations, of Schottky-gated heterostructures based on the highest quality organic single-crystals currently available. As these same devices allow measurements of electron transport down to temperatures ~ 30 K –much lower than what is normally accessed with conventional field-effect transistors- while preserving a high charge carrier mobility. Organic single-crystal Schottky-gated interfaces are not only important for the investigation of organic charge-transfer interfaces, but can also play an important role to explore the intrinsic transport properties of molecular semiconductors[33, 37, 38].

**Acknowledgements**

The authors would like to thank C. Caillier for his assistance during the SKFPM measurements and S. Ono, A. Ferreira, and I. Crassee for assistance. This study was financially supported by MaNEP, the Swiss National Science Foundation, NEDO and the AFOSR (FA9550-08-01-0331).

**Author contributions**

I.G.L. developed and fabricated the devices; performed electrical, Hall and SKFPM measurements; analyzed the data and interpreted the results. M.N. fabricated and characterized the first un-gated rubrene-PDIF-CN$_2$ charge-transfer interfaces. N.A.M. designed the Hall set up and contributed to the electrical characterization. F.V.D.G. contributed to the device fabrication and to the electrical measurements. Z.C. and A.F. synthesized and provided the sample material from which the PDIF-CN$_2$ single crystals where grown. A.F. also contributed to the writing of the manuscript. A.F.M. planned and supervised the work, interpreted the results and wrote the manuscript. All authors contributed to the scientific discussion of the results.

**Additional information**

The authors declare no competing financial interests.

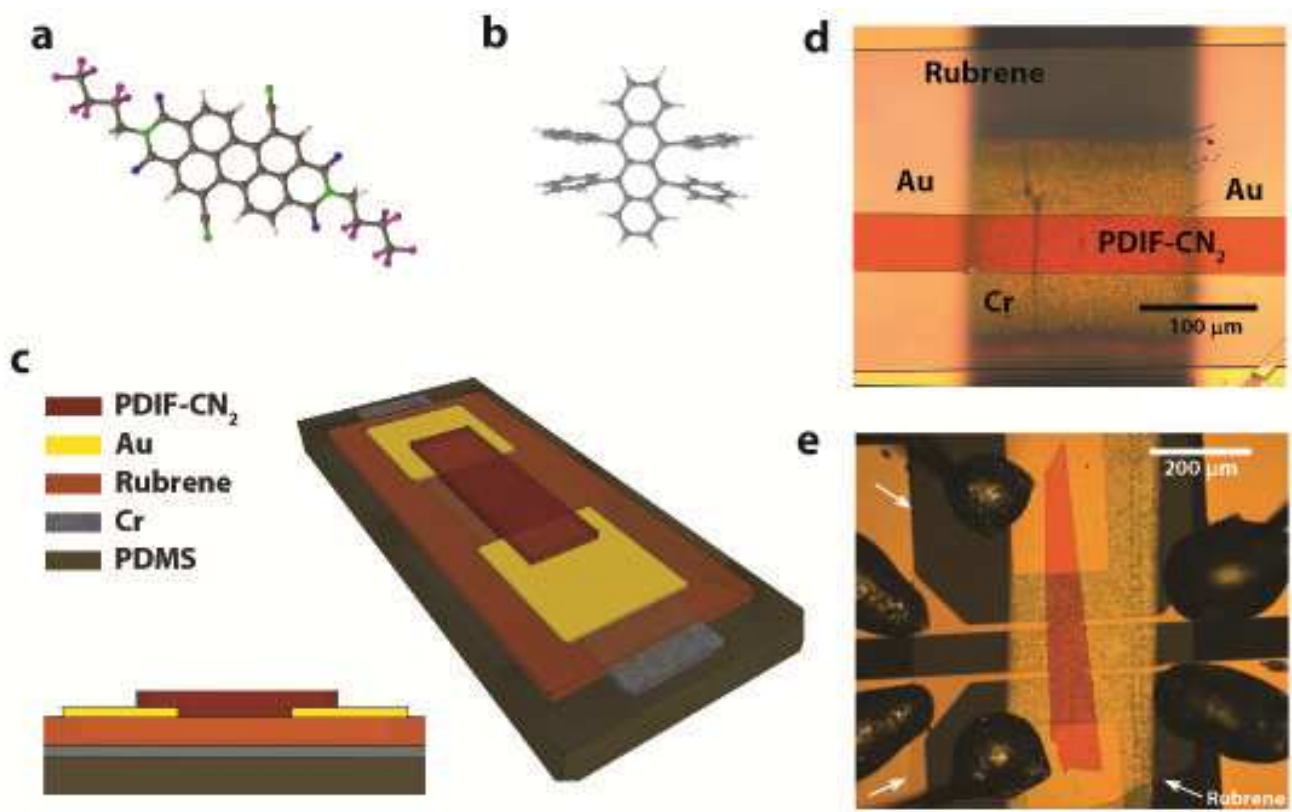

**Fig. 1 Rubrene-PDIF-CN$_2$ Schottky-gated heterostructures.** (a) Chemical structure of the *N,N'*-bis(n-alkyl)-(1,7 and 1,6)-dicyanoperylene-3,4:9,10-bis(dicarboximide) (PDIF-CN$_2$) and (b) rubrene molecule used for the realization of the Schottky-gated organic heterostructures. (c) Schematic representation of the Schottky-gated heterostructures, comprising a PDMS support, the metal gate electrode, a bottom rubrene crystal, the Au source and drain contacts and a top PDIF-CN$_2$ crystal. The interfaces are formed by the high mobility planes of both the rubrene and the PDIF-CN$_2$ crystals. (d-e) Optical microscope images of rubrene-PDIF-CN$_2$ Schottky-gated heterostructures. Panel (d) zooms in on the core of one device and panel (e) shows a larger field image of another (multi-terminal) device that includes the epoxy-bonded contact wires (the bars are 100 and 200 microns in the two images, respectively).

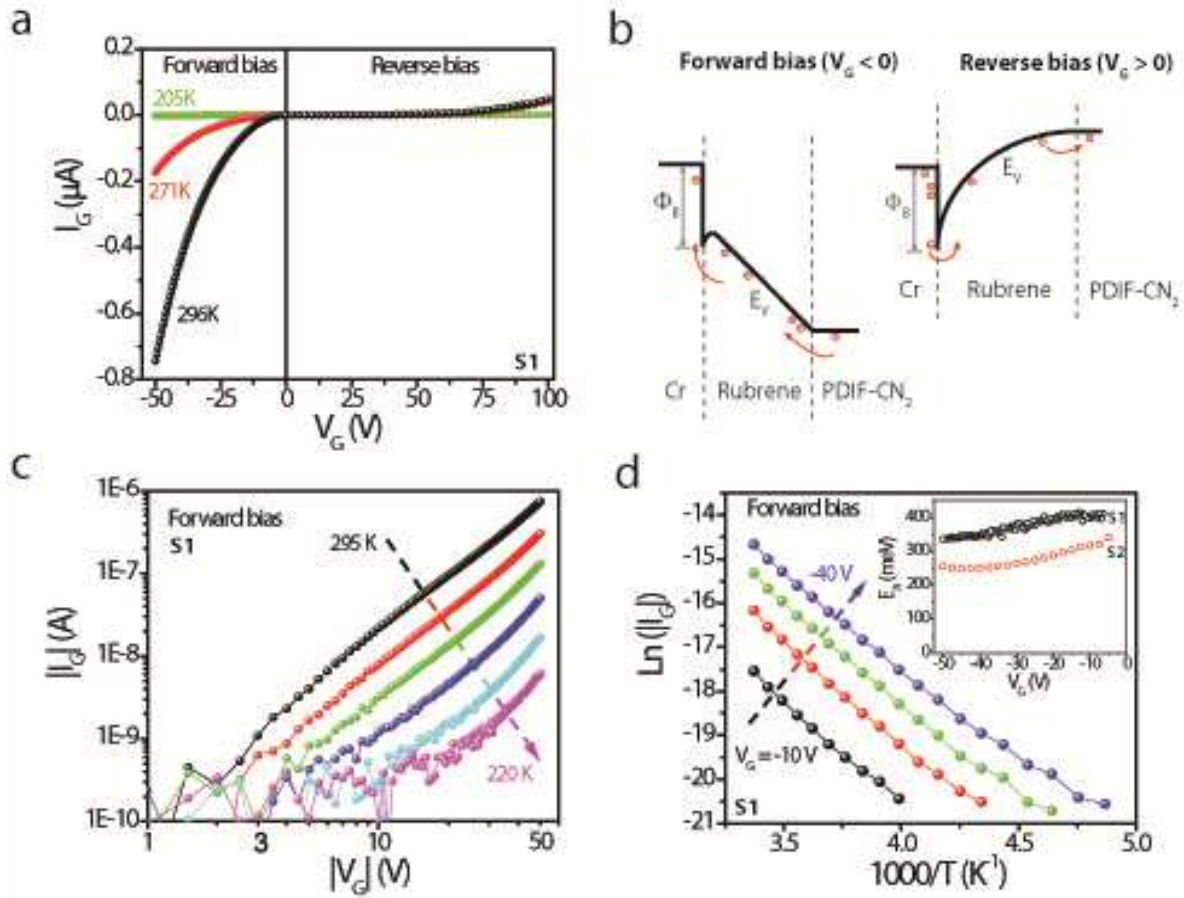

**Fig. 2 Characterization of the Schottky gate.** A careful characterization of the Schottky gate is essential to ensure correct device operation. (a) Current $I_G$ (measured on device S1) flowing through the rubrene crystal upon biasing the gate with respect to the source and drain electrodes. The bias dependence is strongly asymmetric, as expected for a Schottky gate: space-charge limited current flows through the device for negative $V_G$ (forward bias), whereas the current is injection limited by the Schottky barrier at the Cr-rubrene interface for positive $V_G$ (reverse bias; see the schematic energy band diagrams in panel (b)). As the temperature is lowered the current is strongly suppressed also for negative $V_G$, increasing the range of operation where the device is not affected by leakage current. (c) Bias dependence of $I_G$ measured for negative $V_G$ at different temperatures (device S1). The data show a quadratic dependence, characteristic of space-charge limited current. With lowering $T$ the current decreases exponentially, as shown in panel (d), because space-charge limited transport occurs through trap states in the band-gap of undoped rubrene crystals (see also ref. 27). The corresponding activation energy, ranging between 250 and 400 meV (in agreement with ref. 28), is a measure of the energy difference between the Fermi level and the top of the valance band of rubrene (the inset shows data from device S1 and S2).

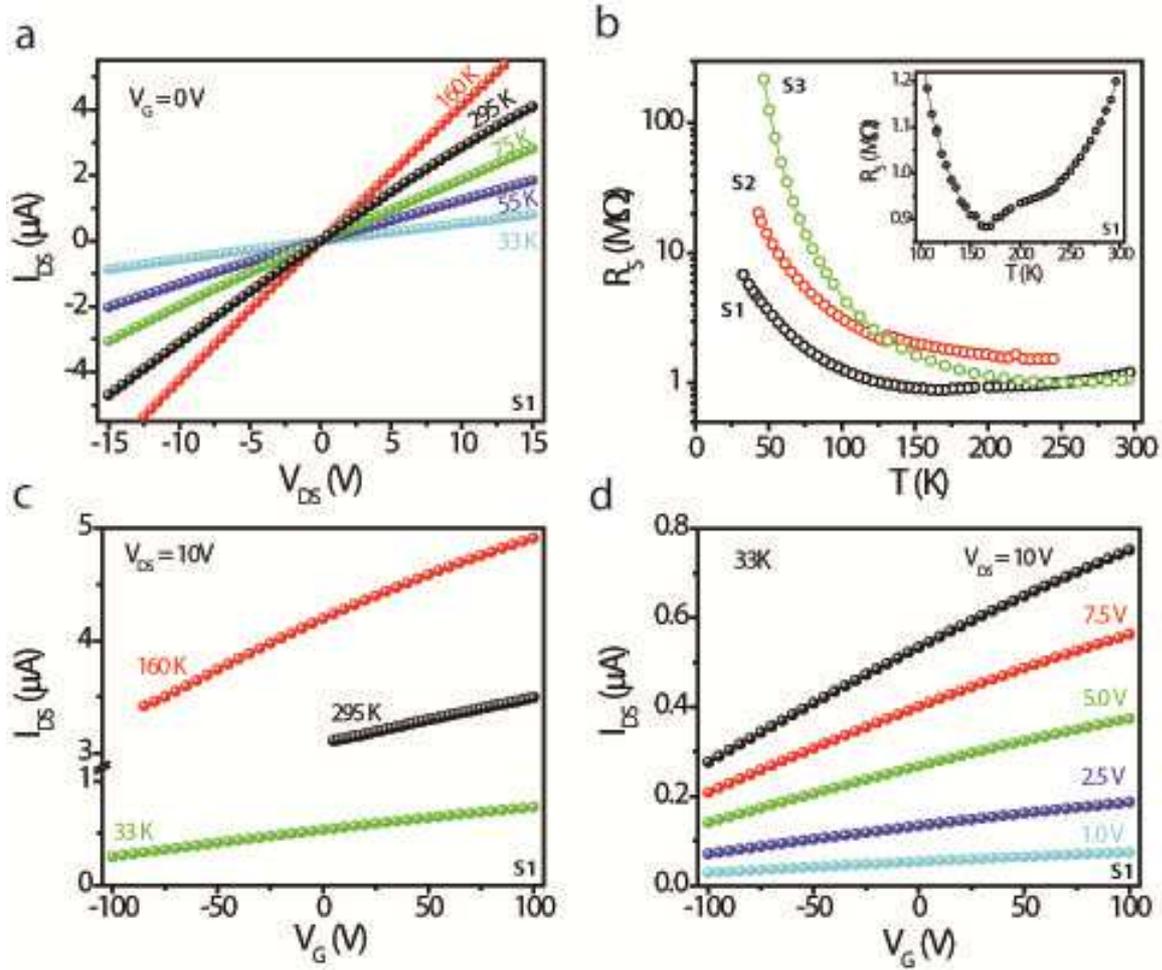

**Fig. 3 In-plane transport characteristics of Schottky-gated rubrene-PDIF-CN$_2$ interfaces.** (a) *I-V* characteristics measured at different temperatures for $V_G = 0$ V. The linearity signals the presence of a conducting layer throughout the investigated temperature range (and indicates the good contact quality). (b) In all cases, the resistivity depends weakly on temperature between room temperature and 100 K, and increases rapidly upon further cooling in all devices. The inset zooms in on the data taken on device S1, showing that in the best samples the resistance decreases with lowering T between room temperature and approximately 150 K. (c) In all devices and at all temperatures the current at the interface increases when a positive gate voltage is applied, indicating that charge carriers are electrons. As the temperature is lowered, the leakage current at negative $V_G$ is strongly suppressed, allowing measurements throughout a larger $V_G$ range (in all the measurements the source-drain current $I_{DS}$ is much larger than the gate leakage current $I_G$) as shown in Fig 2. (d) $V_G$-dependence of the source-drain current at different source-drain voltage, measured at the lowest temperature reached in our experiments (33 K). The labels S1, S2 and S3 indicate on which device the data was measured.

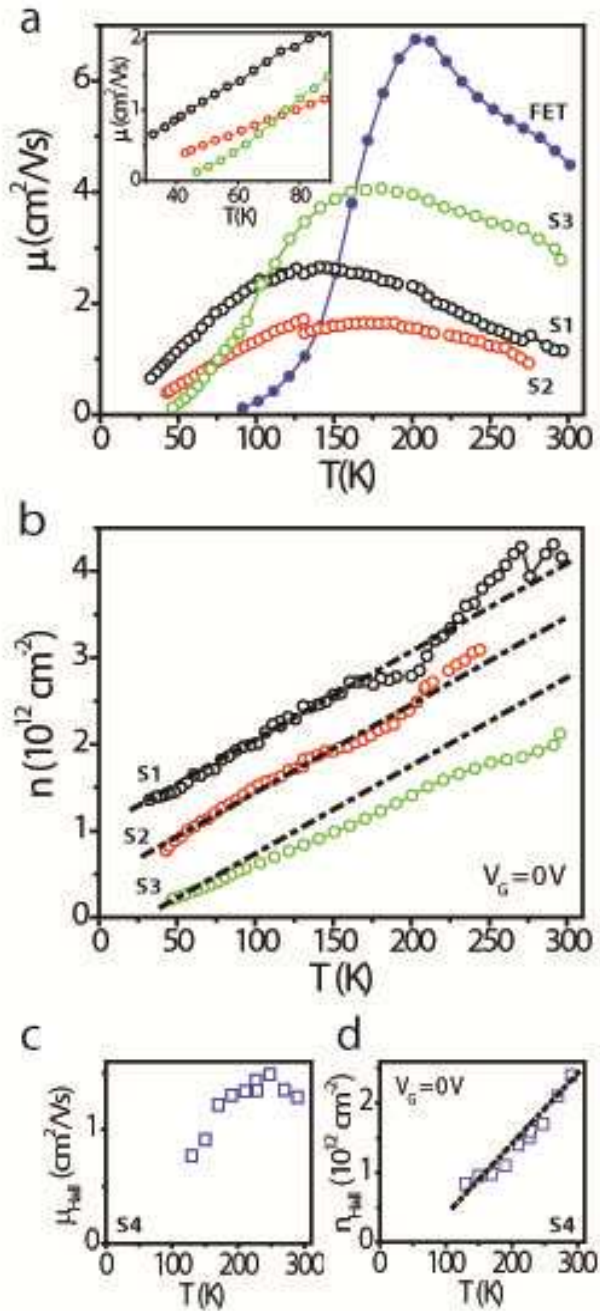

**Fig. 4 Temperature dependence of electron mobility and density at rubrene-PDIF-CN$_2$ interfaces.** (a) The black, red, and green circles represent the field-effect mobility of electrons for heterostructures S1, S2, and S3, respectively (the mobility value are taken at $V_G$ = 0 V; at room temperature, where the applied $V_{DS}$ shifts the onset of the leakage current to positive $V_G$ values, the data is taken at the smallest possible $V_G$). The filled blue circles represent the field-effect mobility measured in a PDIF-CN$_2$ single-crystal FET with vacuum as gate dielectric, which are the *n*-channel transistors with the highest mobility to date. The comparison shows that the heterostructure devices exhibit a band-like transport –increasing mobility with decreasing *T*- in a broader temperature range. In the heterostructures, the mobility remains as high as ~1 cm$^2$V$^{-1}$s$^{-1}$ at the lowest temperature values (see inset), orders of magnitude larger than in the best organic *n*-channel FETs at the same temperatures. (b) Electron density at $V_G$ = 0 V for the same three devices whose mobility data are shown in (a). All devices exhibit the same (within +/- 10%) linear temperature dependence, whose slope matches the prediction of the model calculations discussed in the main text (represented by the dash-dotted lines). (c) and (d) show the temperature dependent electron mobility and electron density (at $V_G$ = 0 V), extracted from Hall measurements performed on device S4, which provide an independent confirmation of the analysis of the field-effect data and of the comparison with the model prediction (dashed-dotted line).

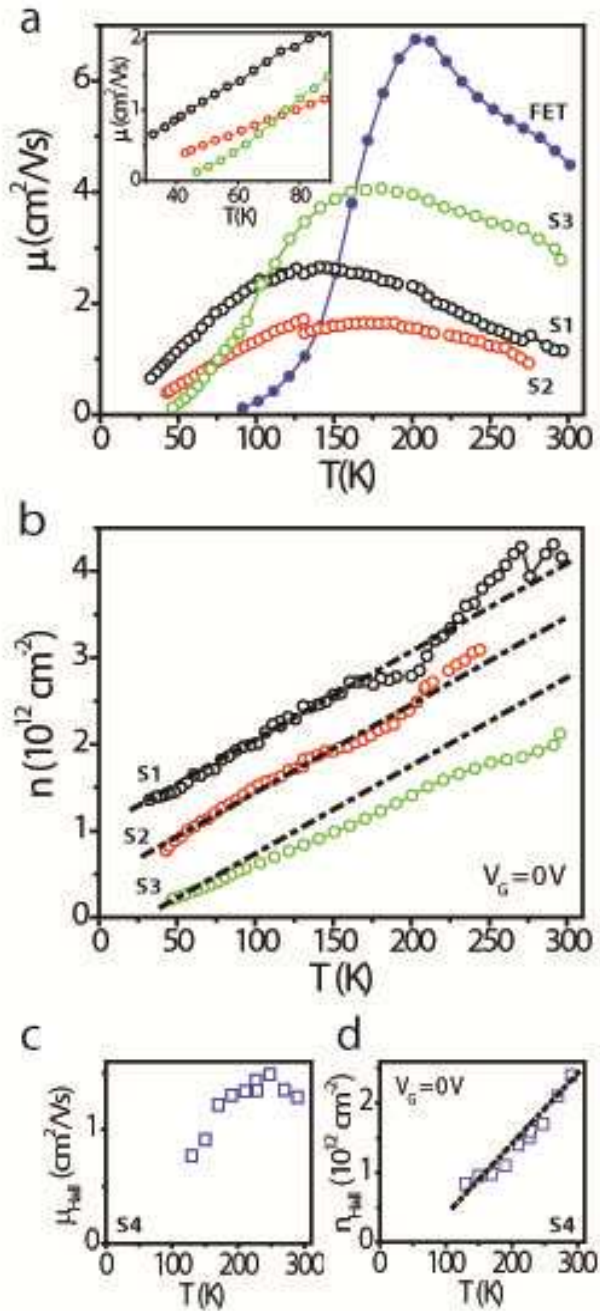

**Fig. 4 Temperature dependence of electron mobility and density at rubrene-PDIF-CN$_2$ interfaces.** (a) The black, red, and green circles represent the field-effect mobility of electrons for heterostructures S1, S2, and S3, respectively (the mobility value are taken at $V_G$ = 0 V; at room temperature, where the applied $V_{DS}$ shifts the onset of the leakage current to positive $V_G$ values, the data is taken at the smallest possible $V_G$). The filled blue circles represent the field-effect mobility measured in a PDIF-CN$_2$ single-crystal FET with vacuum as gate dielectric, which are the *n*-channel transistors with the highest mobility to date. The comparison shows that the heterostructure devices exhibit a band-like transport –increasing mobility with decreasing *T*- in a broader temperature range. In the heterostructures, the mobility remains as high as ~1 cm$^2$V$^{-1}$s$^{-1}$ at the lowest temperature values (see inset), orders of magnitude larger than in the best organic *n*-channel FETs at the same temperatures. (b) Electron density at $V_G$ = 0 V for the same three devices whose mobility data are shown in (a). All devices exhibit the same (within +/- 10%) linear temperature dependence, whose slope matches the prediction of the model calculations discussed in the main text (represented by the dash-dotted lines). (c) and (d) show the temperature dependent electron mobility and electron density (at $V_G$ = 0 V), extracted from Hall measurements performed on device S4, which provide an independent confirmation of the analysis of the field-effect data and of the comparison with the model prediction (dashed-dotted line).

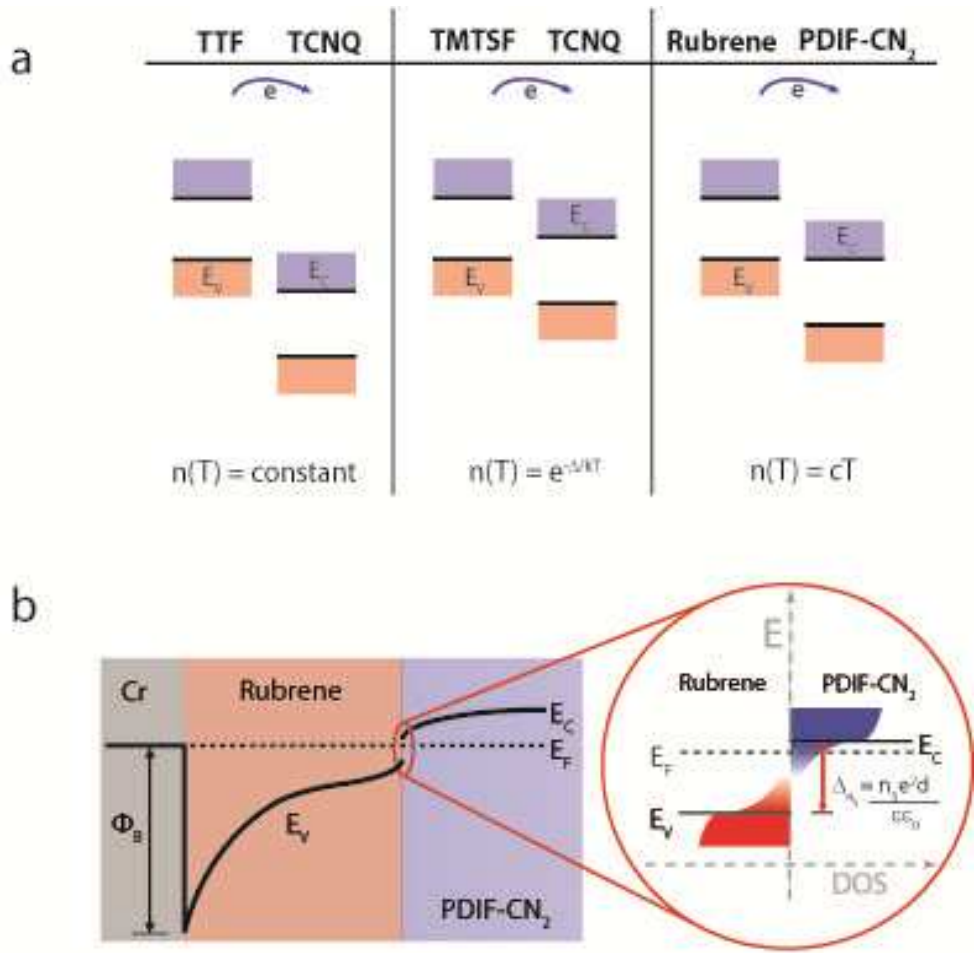

**Fig. 5 Energy level alignment of organic interfaces and band-diagram of rubrene-PDIF-CN$_2$ Schottky-gated heterostructures.** (a) Schematics of the energy-band alignment in organic single-crystal interfaces. TTF-TCNQ (left; see ref. 15) is characterized by a large overlap between the valence band of the electron donor (TTF) and the conduction band of the electron acceptor (TCNQ), resulting in a temperature-independent charge carrier density at the interface. In TMTSF-TCNQ (middle; see ref. 16) a small energy gap between the top of the valence and the bottom of the conduction band of the two materials leads to a thermally activated temperature dependence of the interfacial carrier density. In rubrene-PDIF-CN$_2$ interfaces (right; this work) the top of the valence band of rubrene is nearly perfectly aligned with the bottom of the conduction band of PDIF-CN$_2$, resulting in a linear temperature dependence of the carrier density transferred at the interface. (b) Band diagram of rubrene-PDIF-CN$_2$ Schottky-gated heterostructures (see the supplementary information for details). The small gap at the rubrene-PDIF-CN$_2$ interface originates from the difference in electrostatic (dipole) potential generated by the transferred interfacial charge. The zoom-in of the interfacial region represents the density of states at the surface of rubrene (red) and PDIF-CN$_2$ (blue) as a function of energy. It includes the density of states associated to the shallow traps of the two materials, which decays rapidly away from the valence and conduction band

edges (represented by the two horizontal lines labeled $E_V$ and $E_C$). As discussed in the text, the Fermi level $E_F$ is pinned very close to the bottom of the PDIF-CN$_2$ conduction band.

# Supplementary information

# Single-Crystal Organic Charge-Transfer Interfaces probed using Schottky-Gated Heterostructures

Ignacio Gutiérrez Lezama[1], Masaki Nakano[2], Nikolas A. Minder[1], Zhihua Chen[3, 4], Flavia V. Di Girolamo[1, 5], Antonio Facchetti[3, 4] and Alberto F. Morpurgo[1]

[1] DPMC and GAP. University of Geneva, 24 quai Ernest Ansermet, CH1211 Geneva, Switzerland

[2] CERG, RIKEN Advanced Science Institute, Wako, Saitama 351-0198, Japan

[3] Polyera Corporation, 8045 Lamon Avenue, Skokie, Illinois 60077, USA

[4] CHEM, Northwestern University, 2145 Sheridan Road, Evanston, Illinois 60208, USA

[5] CNR-SPIN and University of Naples, p.le Tecchio 80 80125 Naples, Italy

# 1. Electrical characterization of rubrene and PDIF-CN$_2$ crystals using FETs realized by lamination

The fabrication of Schottky-gated heterostructures relies on the lamination of thin and flat molecular single crystals of rubrene and PDIF-CN$_2$ that are grown using conventional vapor phase transport techniques[S1]. We have extensively investigated the properties of these same crystals by using them for the realization of field-effect transistors, in combination with several different gate dielectrics (for rubrene[S2, S3]: vacuum, parylene, SiO$_2$, Al$_2$O$_3$, Si$_3$N$_4$, and Ta$_2$O$_5$; for PDIF-CN$_2$[S4]: vacuum, Cytop, and PMMA). All these FET devices were also fabricated by means of the same lamination technique. Field-effect transistor measurements are highly reproducible and show that the carrier mobility depends on the dielectric constant of the gate dielectric[S3]. In particular, when vacuum is used as gate dielectric, band-like transport[S5] (i.e., mobility increasing with lowering temperature) is observed both for holes in rubrene (room temperature mobility values up to 20 cm$^2$V$^{-1}$s$^{-1}$ have been reproducibly observed[S5]) and for electrons in PDIF-CN$_2$ (reproducible room-temperature mobility values[S4] up to 5 cm$^2$V$^{-1}$s$^{-1}$). On gate insulators with a relative dielectric constant of 3, comparable to that of organic crystals, the typical mobility values are somewhat lower (approximately 9 cm$^2$V$^{-1}$s$^{-1}$ at room temperature for holes in rubrene with parylene gate insulator[S3]; approximately 2.5 cm$^2$V$^{-1}$s$^{-1}$ for electrons in PDIF-CN$_2$ with PMMA gate insulators[S4]). In these devices an increase in mobility near room temperature –albeit small- is still observed with lowering temperature. In both cases, however the mobility decreases exponentially with lowering temperature below 150 K, to values that are orders of magnitude smaller than 1 cm$^2$V$^{-1}$s$^{-1}$ at temperatures below 50 K. The results obtained for the electron mobility in the Schottky-gated heterostructures discussed in the main text can be directly compared to those of PDIF-CN$_2$ FETs, and from this comparison we find that electrons in Schottky-gated heterostructures exhibit a much more pronounced band-like transport behavior as discussed in the main text.

An explicit comment is worth on the high reproducibility (e.g., the linear temperature dependence of the electron density with the same slope in all devices; see also table S1 for the statistics of the room temperature resistivity, electron density and mobility values) that is observed despite the fact that the effects that are investigated depend sensitively on the properties of an interface that has been assembled under ambient conditions. Even though this reproducibility may seem surprising, devices that we have studied in the past based on organic single crystals and realized by a similar process commonly exhibit such a reproducibility. Past experiments have shown, for instance, an excellent reproducibility of the carrier mobility at rubrene/dielectric interfaces[S2, S3], PDIF-CN$_2$/dielectric interfaces[S4], reproducible low contact resistance at rubrene/nickel interfaces[S6] and reproducible charge transfer at TTF-TCNQ[S7] and TMTSF-TCNQ[S8] interfaces. Other examples from our own work, as well as from the work of other groups, can easily be found. In short, it is by now established that lamination of organic crystals in ambient conditions leads to interfaces exhibiting highly reproducible electronic properties.

## 2. Temperature dependence of the resistance of un-gated rubrene-PDIF-CN$_2$ interfaces

The temperature dependence of the resistance measured in un-gated rubrene-PDIF-CN$_2$ interfaces –shown in Fig. S1- is qualitatively different from that of organic single-crystal charge transfer interfaces investigated previously, based on TTF-TCNQ[S7] or TMTSF-TCNQ[S8]. Specifically, for rubrene-PDIF-CN$_2$ the square resistance is typically 1 MΩ at room temperature and increases only slowly upon lowering temperature down to approximately 150 K (see Fig. S1). In contrast to this, in TTF-TCNQ[S7] the typical square resistance value is ~10 kΩ and the temperature dependence is metallic-like; in TMTSF-TCNQ[S8] the square resistance is 10-30 MΩ and increases exponentially with lowering temperature. These differences are a manifestation of the different band alignment in the different cases, as discussed in the main text.

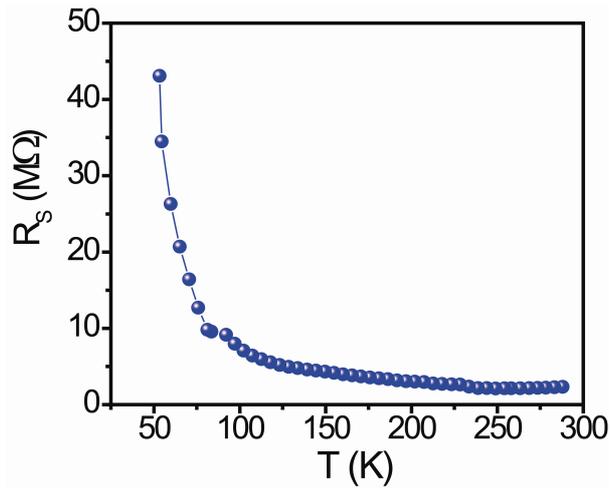

**Figure S1.** Temperature dependence of the square resistance of an un-gated rubrene-PDIF-CN$_2$ interface, showing the weak temperature dependence between room temperature and 100 K that we referred to in the main text.

## 3. Fabrication of Schottky-gated heterostructures and characterization

The rubrene-PDIF-CN$_2$ heterostructures are mounted on a PDMS stamp support, according to the following procedure: We first evaporate through a shadow mask a gate electrode consisting of a strip (approximately 200 or 400 micron wide) of Ti (3 nm, adhesion layer), Au (20 nm, to improve the mechanical properties), Ti (3 nm, to avoid gold interdiffusion), and Cr (20 nm). Although simpler structures may be possible as well, we found that this layer sequence eliminates most problems associated to the mechanical properties and electrical continuity of the gate electrode. Cr was chosen because –with a work function of approximately 4.5 eV[i] – it leads to a sufficiently high Schottky barrier at the interface with rubrene to enable Schottky gating, while being a simpler material to process and handle as compared to metals with a smaller work function. A rubrene crystal (approximately 1-2 micron thick) is subsequently manually aligned under an optical microscope and laminated on top of the gate. Gold contacts (15 nm thick) are then evaporated through a shadow mask directly on top of rubrene. Finally, a thin PDIF-CN$_2$ crystal is laminated onto the rubrene one, so that it makes contact with the gold electrodes and it is aligned with the underlying gate. Small (25 micron in diameter) gold wires are attached to the gold electrodes and to the gate, using solvent-free silver epoxy. Images of devices realized in this way are shown in Fig. 1 of the main text.

While being conceptually similar to the process that we commonly use for the realization[S9] of single-crystal transistors, the fabrication of heterostructures is considerably more complicated in practice, because it requires more steps and rather precise manual alignment. As a consequence, it much more frequently happens (as compared to the fabrication of single-crystal FETs) that a device is damaged during the fabrication (crystal cracking and fracturing of the gate electrode are two of the more commonly encountered problems). For this reason we carefully checked devices before starting their in-depth study, and found that it is particularly important to ensure the electrical continuity of the gate electrode and that the leakage current to the gate is sufficiently small (the leakage current can be strongly increased if the devices are damaged during the fabrication).

Transport measurements through devices that meet the above standards were mainly performed in a two terminal configuration. We found that the resistivity of the interfacial plane in all the measured devices is quite reproducible, with a value ~ 1 MΩ/□ (see table S1 for statistics on all the properly functioning devices). Table S1 also includes the values of carrier density and field-effect mobility for all devices in which these quantities could be extracted.

---

[i] See, for instance, ref. 23 in the main text. It is certainly possible that the work function of our films is somewhat different from this literature value. However, none of our conclusions depends on the precise value of the work function of Chromium, and the only essential issue is that the work function is sufficiently small to allow an effective operation of the Schottky gate.

| No. | Resistivity (MΩ) | n ($10^{12}$ cm$^{-2}$) | µ (cm$^2$/Vs) |
|---|---|---|---|
| 1 | 1.00 | 2.27 | 2.75 |
| 2 | 1.20 | 4.16 | 1.15 |
| 3 | 1.20 | 4.33 | 1.20 |
| 4 | 1.25 | 6.24 | 0.80 |
| 5 | 1.28 | 3.48 | 1.40 |
| 6 | 1.37 | Not gated ||
| 7 | 1.46 | 2.19 | 1.95 |
| 8 | 1.50 | Device broke ||
| 9 | 1.60 | 1.70 | 2.28 |
| 10 | 1.63 | Not gated ||
| 11 | 2.00 | 4.46 | 0.70 |
| 12 | 2.06 | 2.80 | 1.10 |
| 13 | 2.20 | 3.05 | 0.93 |
| 14 | 2.27 | 0.91 | 3.0 |
| 15 | 2.48 | Not gated ||

**Table S1**. Resistivity, electron density and electron mobility extracted from two-terminal measurements. These results include all the functioning devices measured at room temperature.

In this work we focused mainly on the investigation of transport in a two-terminal configuration, because of the complexity of making four-probe or Hall bar devices, in which the PDIF-CN$_2$ crystals have to be aligned with 4 or 6 contacts instead of two. This implies that the last alignment step needs to be carried out with much better precision, which requires lengthier manipulations of the crystals and easily results in damage and in mechanical stress which can reduce the carrier mobility. We found that this type of degradation poses particularly serious challenges to the fabrication of Hall bar devices, where alignment of the PDIF-CN$_2$ crystal requires a precision of ~ 10 μm. We also found that the presence of additional contacts renders the devices more susceptible to cracks that appear in the crystals as devices are cooled down. For all these reasons, basing our investigations on multi-probe devices would have very seriously limited the amount of properly functioning devices that we could have studied in detail. We nevertheless succeeded in fabricating working devices with four or six probes. In these devices we could compare the square resistance extracted from two- and four-probe measurements, and we found no significant difference (even at lower temperature), as shown in Fig. S2, which illustrates the comparison between the two and four-terminal resistivity in two different devices. Fig. S2a corresponds to device S3 whose mobility and electron density are shown in Fig 4a and 4b respectively. The data in Fig S2b corresponds to device S4, on which the Hall measurements where performed and which are described in the next section. These measurements demonstrate that the contact resistance is small and does not significantly influence the results of this work.

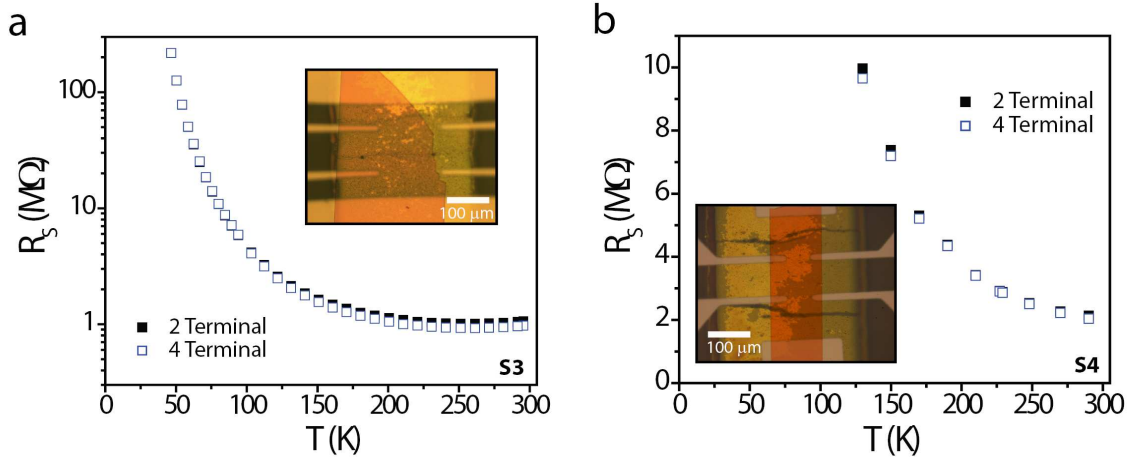

**Figure S2.** Resistivity (at $V_G = 0$ V) of two heterostructures as a function of temperature, measured in a two- terminal (full squares) and in a four-terminal (open squares) configuration. The comparison shows that the contact resistance has a negligible contribution to the total resistivity of the heterostructures. a) corresponds to one of the devices whose two-terminal resistivity, electron density and mobility are discussed in the main text, while b) corresponds to the resistivity of the device in which the Hall effect was measured. The insets show an optical image of the devices.

## 4. Hall measurements on Schottky-gated heterostructures

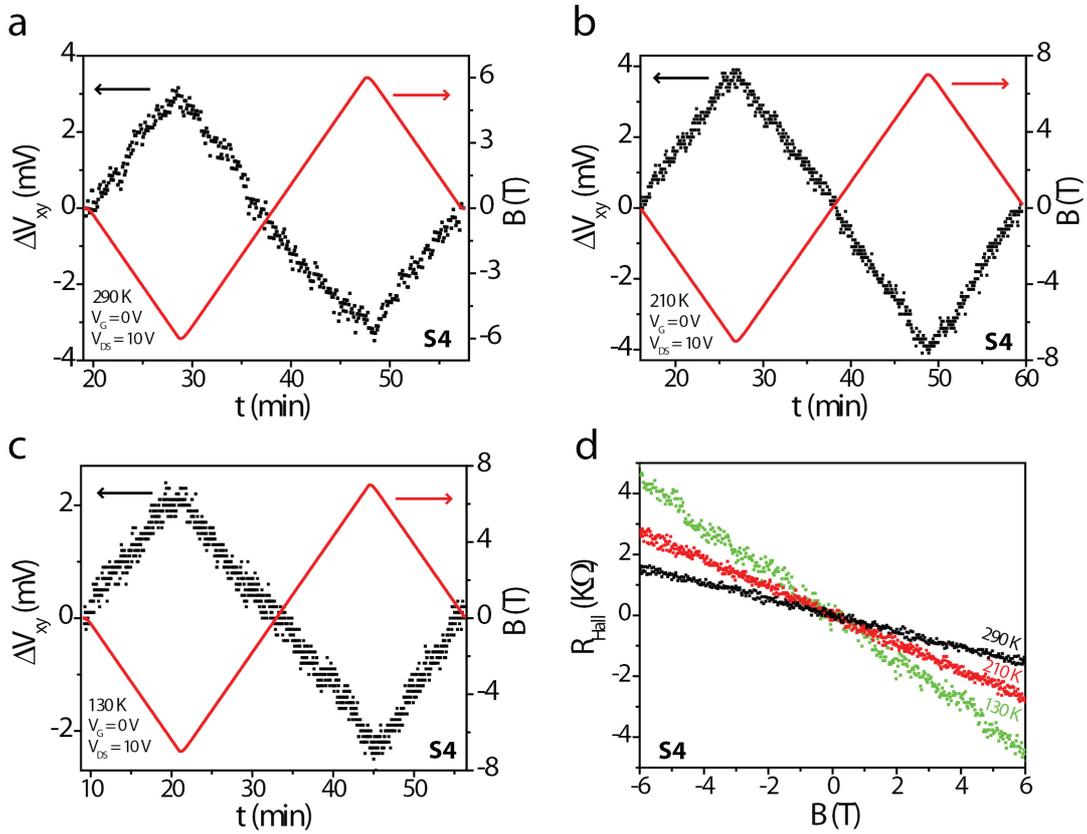

**Figure S3.** Time evolution of the Hall voltage $\Delta V_{xy}$ as a function of magnetic field, at a) 290 K, b) 210 K, and c) 130 K. A voltage offset at $B = 0$ T (about 8% of the applied $V_{DS}$) has been subtracted. d) shows the Hall resistance as a function of B. The negative slope indicates electron transport.

We succeeded in measuring the Hall effect as a function of temperature on one device, and in a few other devices Hall effect measurements could also be performed, but only close to room temperature (as the devices broke upon cooling). These experiments corroborate the results obtained by FET measurements, as they provide an independent measurement of the type of charge carriers contributing to transport, and of the temperature dependence of their carrier density and mobility.

The measurements were performed in a variable temperature cryostat, in a perpendicular magnetic field up to 7 T, and sweeping rates up to 0.6 T/min. We focused on measurements of the Hall effect at $V_G = 0$ V: the Hall voltage to be measured is very small and its measurement requires very stable devices (stability of better than one part in one thousand over a few hours) and we found that the application of a gate voltage was not compatible with such a stability level (mainly because of small, slow drifts in the source drain current). The time evolution of the Hall voltage $\Delta V_{xy}$ while sweeping the magnetic field, measured at three different temperatures is shown in Fig. S3a-c. The devices are voltage biased ($V_{DS} = 10$ V) and the source drain current $I_{DS}$ decreases with lowering temperature (in this device the resistance increases with lowering temperature, see Fig. S2b), which makes the Hall signal more difficult to measure at low temperature. Nevertheless, as it can be seen from the raw data, the signal to noise ratio remains excellent even at the lowest temperatures.

From the Hall voltage we extract the Hall resistance (see Figure S3d), whose slope as a function of $B$ increases upon lowering temperature, indicating that the electron density decreases as expected. Indeed, the results shown in Fig. 4 of the main text show that the decrease in electron density is linear with T, and that its slope equals the one obtained from the FET analysis (i.e., it coincides with the one predicted by the model discussed in the main text).

## 5. Scanning Kelvin probe force microscopy across rubrene-PDIF-CN$_2$ heterostructures

Scanning Kelvin probe force microscopy (SKPFM) is a non-contact scanning probe technique commonly used to measure the work function of metals[S10] and semiconductors[S11]. In this technique, a conductive atomic force microscope tip is scanned above a grounded sample, and is capacitively coupled to it. The work function difference between tip and sample results in a contact potential difference (CPD) between the two, which causes an electrostatic force acting on the tip. The SKPFM measurement consists in zeroing this electrostatic force by applying an external voltage that equals the CPD: the value of this voltage corresponds to the work function difference between tip and sample. If the work function of the tip is known, one directly obtains the work function of the sample (for more details see ref. S10). SKPFM has been widely applied to organic semiconductors[S12], for instance to measure their work function[S13], to probe the electrostatic potential profile of field-effect transistor under biased conditions[S14], and to investigate the influence of illumination on the electrostatic profiles at organic interfaces in solar cell structures[S15].

The measurements on rubrene/PDIF-CN$_2$ interfaces were performed in ambient conditions, with an Asylum Cypher scanning force microscope, in a two-scan mode. In the first scan, the topography is measured and during the second scan the measured height profile is used to maintain a constant distance between tip and sample, while measuring the CPD as a function of position. We performed SKPFM measurements on rubrene/PDIF-CN$_2$ interfaces, on actual devices (see Fig. S4a) and on interfaces assembled on different substrates. As an example, Fig. S4b and Fig. S4c show the topography and CPD measured in the area delimited by the red-line in Fig. S4a. On the left side of the image, the surface of the rubrene crystal is exposed and on the right side, the rubrene crystal is covered by a PDIF-CN$_2$ crystal. The height difference between the two –corresponding to the thickness of the PDIF-CN$_2$ crystal- is clearly apparent in Fig. S4b. Fig. S4c shows that in correspondence of the height step, a CPD step is present. For a quantitative analysis, Fig S4d shows the averaged height and CPD step perpendicular to the interface, over the entire area of the images. The difference in CPD on the two materials gives their work function difference irrespective of the absolute value of the work function of the tip (which is an important point to achieve high accuracy, because uncertainties in the work function of the tip can give large calibration errors in absolute measurements). We performed measurements on more than 10 interfaces, using two different type of tips (i.e. with different work functions) and found that the work function difference between rubrene and PDIF-CN$_2$ single crystals is 410 meV +/- 50 meV.

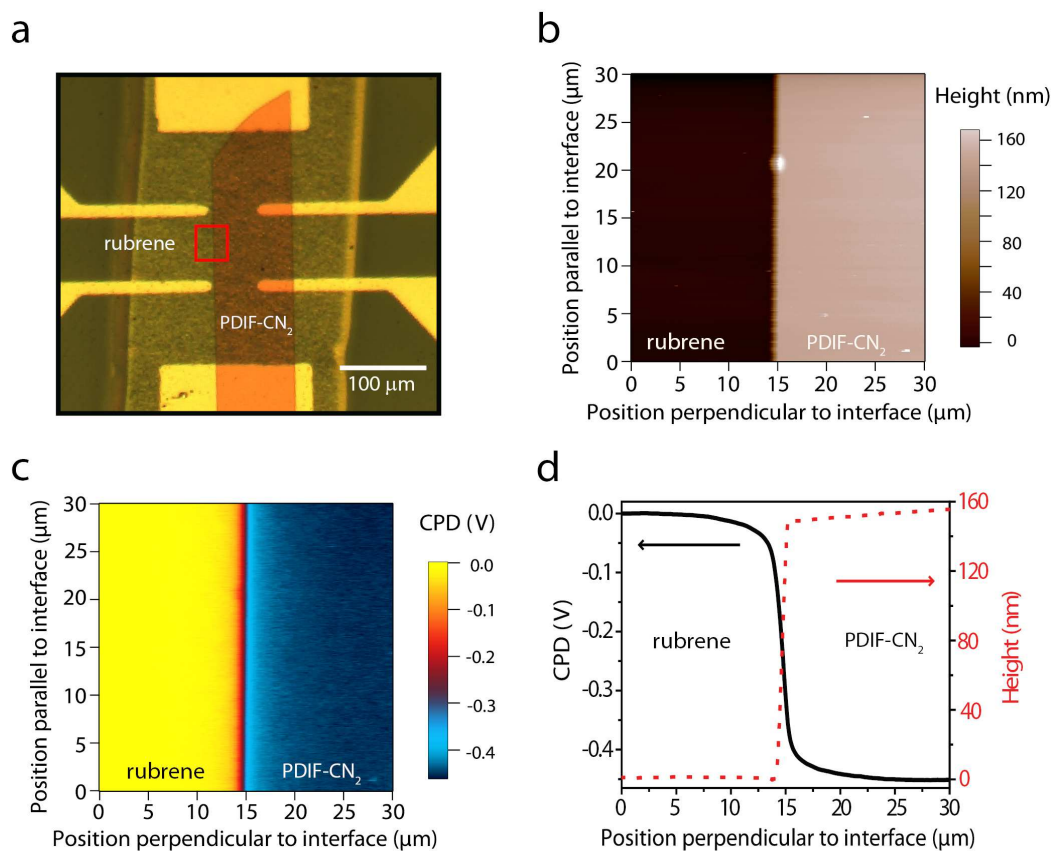

**Figure S**4. Scanning Kelvin force probe microscopy measurements on rubrene-PDIF-CN$_2$ heterostructures. a) shows the optical image of a device. The red rectangle delimits the area in which the measurements shown in the other panels are taken. b) shows the topographic image: a step of 160 nm corresponding to the thickness of the PDIF-CN$_2$ crystal is clearly visible. c) Shows the corresponding CPD measured at in the same region. In d) both the CPD and height profile average (perpendicular to the interface) are shown. A CPD step is observed as the tip is scanned across the interface, from which we can read directly the work function difference between rubrene and PDIF-CN$_2$ crystals.

Together with transport measurements performed on individual rubrene and PDIF-CN$_2$ crystals, the work function difference measured by SKPFM can be used to verify the level alignment between the valence band of rubrene and the conduction band on PDIF-CN$_2$ crystals. As mentioned in the main text, we know from transport measurements (on individual rubrene and PDIF-CN$_2$ crystals) that the energy difference between the Fermi level and the top of the valence band in rubrene is about 300 meV (with sample-to-sample fluctuations of about 100 meV –see inset of Fig. 2d), while the energy difference between the Fermi level and the bottom of the conduction band of PDIF-CN$_2$ is about 100 - 120 meV. Hence if the valence band of rubrene and the conduction band of PDIF-CN$_2$ are aligned, one would expect a work function difference of about 400 meV, with an uncertainty of about 100 meV. This finding is in excellent agreement with the work function difference found via SKPFM measurements, and provides an independent quantitative confirmation of the alignment of the level in rubrene and PDIF-CN$_2$ which was assumed to model the interfacial transport data (see main text and next section).

## 6. Interfacial charge transfer and band diagram of Rubrene-PDIF-CN$_2$ heterostructures

Here we analyze in more detail the band diagram that determines the electronic properties of rubrene-PDIF-CN$_2$ Schottky-gated heterostructures, as well as the two different contributions to the density of electrons present at the rubrene-PDIF-CN$_2$ interface. We start by discussing the interface between semi-infinite rubrene and PDIF-CN$_2$ crystals, and see later what is the effect of the presence of a Schottky gate at a finite distance from the rubrene-PDIF-CN$_2$ interface.

As mentioned in the main text, our experimental observations indicate the absence of both any sizable gap or band overlap between the top of the valence band of rubrene and the bottom of the conduction band of PDIF-CN$_2$, i.e., before bringing the material into contact, the top of the valence band of rubrene is aligned with the bottom of the conduction band of PDIF-CN$_2$. The discussion of the entire band-diagram also requires the –at least approximate– knowledge of the position of the chemical potential in the bulk of the two semiconductors. As we discussed in the main text, the chemical potential is approximately 300 meV above the top of the valence band in rubrene (see ref. 28 and the considerations in main the text about the activation energy of the space charge limited current that flows through the rubrene crystal upon biasing the Schottky gate) and 100 meV below the bottom of the conduction band of PDIF-CN$_2$ (as determined from the activation energy of the residual conductivity measured in individual PDIF-CN$_2$ crystals). Before bringing the two materials into contact, therefore, the relative position of the levels and chemical potentials in the two materials is as depicted in the left side of Fig. S5[ii]

Upon bringing the two materials into contact, electrons redistribute to create a uniform electrochemical potential through the entire structure. There are two contributions to the charge redistribution: a surface contribution, and a contribution due to the displacement of charge in the bulk of the semiconductors potential. The surface contribution is confined to a layer of the thickness of one (or at most few) molecules on each side of the interface, the same thickness that characterizes the accumulation layer in organic field-effect transistors. This contribution involves an equal number of electrons at the surface of PDIF-CN$_2$ and holes at the surface of rubrene, because electrons are excited from the valence band of rubrene to the conduction band of PDIF-CN$_2$. It generates an electrostatic dipole that shifts apart the top of the valence band in rubrene and the bottom of the conduction band in PDIF-CN$_2$, effectively opening a gap proportional to the density of transferred surface charge. The bulk contribution, on the contrary, displaces electrons from a deeper region of the rubrene crystal to the PDIF-CN$_2$ crystal. It is driven by the chemical potential in the bulk of the rubrene crystal, which is higher than that in the bulk of PDIF-CN$_2$. The characteristic length scale for this contribution to the charge redistribution is the usual screening length of doped semiconductors, which for

---

[ii] We are discussing the electronic properties of organic semiconductors in full analogy with those of conventional semiconductors, how it is often done. In this context, we consider that the position –in energy- of the valence and conduction bands in the two organic crystals also includes terms such as the re-organization energy, which are relevant for organic materials but not for inorganic ones. Additionally, correlation energies due to electron-electron interactions –that could be important in organic semiconductors owing to their rather narrow band-width- are neglected because the density of interfacial charge carriers is about 0.01 per molecule (much smaller than 1 per molecule) .

the small doping levels typical of high-quality organic crystals corresponds to several hundreds of nanometers. The final band diagram therefore is the one depicted on the right of Fig. S5.

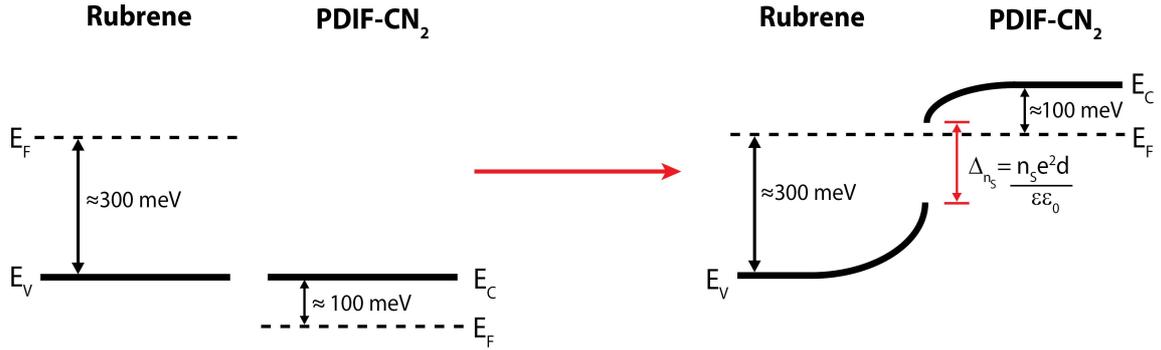

**Figure S5** Left: alignment of the bands in rubrene and PDIF-CN$_2$, and position of the chemical potentials in the two materials before bringing them into contact. Right: band-diagram in the interface region after establishing contact between the two materials. The gap between the top of the valence band of rubrene and the bottom of the conduction band of PDIF-CN$_2$ is due to the surface charge density n$_s$ transferred between the two materials. Band bending in the bulk, away from the interface occurs on a length scale determined by the doping level (typically few hundreds

The surface contribution to the charge density can be calculated by simply looking at the probability of exciting an electron from the rubrene to the PDIF-CN$_2$ surface. If a small gap $\Delta$ was present between the top of the rubrene valence band and the bottom of the PDIF-CN$_2$ conduction band, the situation would be identical to that of TMTSF-TCNQ interfaces, which we have discussed in ref. S8. The corresponding expression for the density of transferred charge reads:

$$n_S = \int_{\Delta/2}^{\infty} 2 N_S e^{-\frac{E}{kT}} dE \tag{S1}$$

where $N_S$ is the density of states per unit surface (which is approximately the same for rubrene and PDIF-CN$_2$, and is given by the density of molecules -1 x 10$^{15}$ cm$^{-2}$- divided by the bandwidth, ~0.5 eV; the factor 2 in the expression comes from spin degeneracy). Even though for rubrene-PDIF-CN$_2$ no gap is present before charge transfer (i.e. $\Delta = 0$ before bringing the two semiconductors together) a gap is effectively opened by the electrostatic dipole created by the surface charge transfer, once the two semiconductors are brought into contact. From simple electrostatics, the expression for this gap (i.e., electrostatic potential associated to the dipole) is:

$$\Delta = \frac{n_S e}{\varepsilon \varepsilon_0} ed \tag{S2}$$

Here, $d$ is the distance between the electron and hole layers at the two surfaces (layer which we take to be 1 nm) and $\varepsilon$ is the relative dielectric constant of the organic materials (which we take to be 3). We therefore get a self-consistent equation for $n_S$:

$$n_S = \int_{\frac{n_S e^2 d}{2\varepsilon \varepsilon_0}}^{\infty} 2 N_S e^{-\frac{E}{kT}} dE = 2 N_S kT e^{-\frac{n_S e^2 d}{2\varepsilon \varepsilon_0 kT}} \tag{S3}$$

which is equation (1) in the main text.

Charge transfer from the rubrene to the PDIF-CN$_2$ surfaces generates as many electrons on the PDIF-CN$_2$ surface as holes on the rubrene surface. However, the alignment of the electrochemical potential in the bulk of the two crystals gives an additional electron contribution (see Fig. S5) that fills the hole states, which is why only electron transport is observed. The total electron density at the interface is then determined by the values of the chemical potentials in the two materials, which, in turn, is determined by the amount of unintentional dopants present and by the disorder-induced density of states inside the band-gap of the two semiconductors, and that therefore fluctuates from sample to sample. This is why the total electron density that we find in different devices is different.

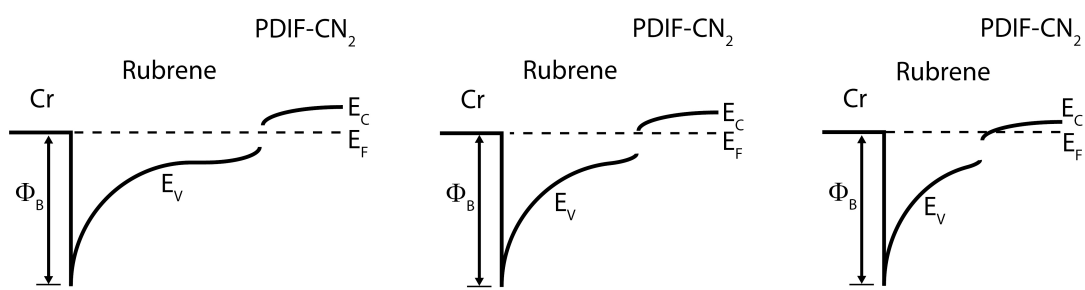

**Figure S6** Band diagram of Schottky-gated rubrene-PDIF-CN$_2$ heterostructures for different thicknesses of the rubrene crystal. Left: the rubrene crystal is thicker than the depletion region associated to the Schottky barrier. The band bending due to the Schottky barrier formation does not overlap with the band bending due to the charge accumulated at the interface (as indicated by a flat band region in between the two) and can be treated independently. Middle: if some overlap is present between the depletion region and the charge accumulation region at the rubrene-PDIF-CN$_2$ interface, the net result is to shift the Fermi level at the interface closer to the bottom of the conduction band in PDIF-CN$_2$ (as can be easily understood by considering the case –on the right- in which the interface is much closer to the Schottky gate). Indeed it may be possible to accumulate rather high electron densities at the interface in devices with a sufficiently thin rubrene crystal (although very interesting, these devices are harder to fabricate and are expected to exhibit higher leakage current at room temperature, even though at lower temperature the leakage current may still be completely suppressed).

It is now simple to understand the effect of the finite thickness of the rubrene crystal, and of the Schottky barrier formation. The Schottky barrier at the Cr/rubrene interface leads to the formation of the usual depletion region. If the rubrene-PDIF-CN$_2$ interface is further away from the Cr gate electrode (i.e., if the thickness of the rubrene crystal is much larger than the depletion region associated to the formation of the Schottky barrier), the band bending due to the Schottky barrier itself and to the rubrene-PDIF-CN$_2$ interface are spatially separated (see left scheme in Fig. S6) and can be treated independently. In practice, the depletion region can extend up to nearly 1 micron (similarly to what has recently been found for rubrene metal-semiconductor transistors[S16]), at least for the devices in which the concentration of dopants is lowest. In this case the depletion region due to the Schottky barrier can have some overlap with the region (in the rubrene crystal) where band bending originating from the rubrene-PDIF-CN$_2$ interface is present (see middle scheme in Fig. S6). The result of this overlap is to further shift upward the electrochemical potential at the interface (as it can be simply understood by considering the case of a thin rubrene crystal, see right scheme in Fig. S6), and provides an additional contribution to the electron density at the interface. The resulting complete band diagram of the Schottky gate heterostructure is shown in Fig. 5.